\documentclass[twocolumn,showpacs,preprintnumbers,amsmath,amssymb,prb]{revtex4-1}

\usepackage{graphicx}
\usepackage{dcolumn}
\usepackage{bm}
\def\3he{$^3$He}
\def\4he{$^4$He}

\begin{document}


\title{Growth of solid hcp \4he off the melting curve}

\author{M.W. Ray}
\author{R.B. Hallock}%
\affiliation{%
Laboratory for Low Temperature Physics, Department of Physics,\\
University of Massachusetts, Amherst, MA 01003
}%

\date{\today}

\begin{abstract}
We report studies of the growth of solid hcp \4he at pressures higher than the bulk freezing pressure using a cell design that allows us to inject atoms into the solid. Near the melting curve during injection we observe random events during which the pressure recorded in the cell drops abruptly. These events are accompanied by transient increases in the temperature of the cell. We discuss these transients and conclude that they represent the solidification of meta-stable liquid regions and the associated relief of strain in the local solid. We also observe that further from the melting curve the transients are no longer recorded, but that we can continue to add atoms to the solid, increasing its density at fixed volume.  We document these changes in density with respect to changes in the chemical potential as a function of temperature and discuss these in the context of recent theoretical work.  
\end{abstract}

\maketitle

\section{Introduction}

Supersolidity, which was first predicted almost 50 years ago \cite{Andreev1969,Chester1970,Leggett1970}, has received substantial attention during the past several years.  Stimulated by the work of Ho, Bindloss and Goodkind\cite{Ho1997}, Kim and Chan \cite{Kim2004a,
Kim2004b, Kim2005} carried out torsional oscillator experiments in which an anomalous shift in the resonant period of a torsional oscillator filled with solid \4he was observed below about 250 mK.  This was interpreted as evidence for a supersolid phase in solid \4he. Although the interpretation of supersolidity is still controversial, it is now believed that this period shift, or non-classical rotational inertia (NCRI), may have its origin in disorder
in the crystal \cite{Balibar2008}.  To date the bulk of the evidence for unusual behavior comes from torsional oscillator measurements, although experiments of the shear modulus have shown  unexpected behavior \cite{Day2007,Day2010,Rojas2010} in the same range of temperature.

If solid \4he is a supersolid, it would be expected to support mass flow.  But, attempts to observe such flow by directly squeezing the solid\cite{Greywall1977,Day2005,Day2006,Rittner2009} have found no evidence for flow.  By use of a conceptually different approach, we previously reported on the observation of mass flow through solid \4he \cite{Ray2008a,Ray2009,Ray2009b,Ray2010} during experiments in which a chemical potential difference was created across the solid by directly injecting mass into one side of the solid.  The mass flow was only observed when the temperature of the solid was less than $\lesssim 550$ mK.  It was further noted that when flow was observed the pressure of the solid changed in the fixed-volume cell, but when no flow was observed, no change in the pressure in the solid was recorded.  Finally, with two pressure gauges mounted on our cell, we also noted that stable pressure differences were often present across the solid. These pressure differences had no bearing on whether or not flow was observed.  These observations have led us to study, in more detail, the growth of solid helium from the superfluid.

We have employed the cell designed for our mass flow experiments \cite{Ray2008a,Ray2009,Ray2009b,Ray2010} to study
the growth of solid helium at pressures greater than the bulk freezing pressure, $P_F$. The design exploits the properties of liquid helium in Vycor, which is a porous glass with a characteristic pore diameter $\sim$7 nm. It is well known that inside the Vycor at low temperatures the melting curve is elevated to $P_V \approx 37$ bar \cite{Beamish1983,Lie-zhao1986,Adams1987}.  This elevation of the melting curve allows us to have an interface between superfluid (in the Vycor) and the bulk solid (in the cell) so we can readily create a chemical potential difference between our fill lines, or between them and the solid.

We study the effect of injecting mass into the solid and show that the solid can grow at constant volume when $P > P_F$.  We present two central results:  (1) In the immediate vicinity of the melting curve we see transients in the temperature of $\sim 10$ mK accompanied by pressure drops which can be up to $~100$ mbar.  We discuss the possibility that these transients are due to the solidification of meta-stable liquid regions.  (2) Further from the melting curve we no longer observe these transients but we are still able to grow the solid.  We discuss this isochoric compressibility in the context of the theory of the superclimb of edge dislocations \cite{Soyler2009}.  Also, we will comment further on the appearance of pressure differences seen across the solid.

\begin{figure}
\resizebox{2 in}{!}{
\includegraphics{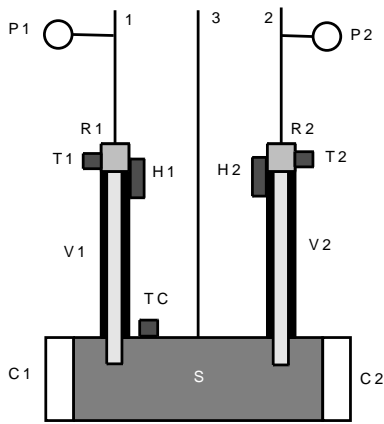}}
\caption{\label{fig:cell} Cell used to study the growth of solid
helium from the superfluid.  Helium is admitted to the solid
chamber S through capillaries 1 and 2 (heat-sunk only at 4 K)
which first lead to liquid reservoirs atop thin Vycor rods V1 and
V2.  The reservoirs are heated by heaters H1 and H2.  Two
capacitance strain gages, one on each side of S measure the
pressure of solid; the temperature is measured by a calibrated carbon
thermometer TC.  The pressures of the fill lines are measured by
pressure transducers P1 and P2 located outside the cryostat. A
third capillary, 3, heat sunk in several places including the
coldest heat exchanger (but not the mixing chamber), bypassed the
Vycor and was used to initially fill the cell with helium. }
\end{figure}

\section{Experiments and Discussion}
\subsection{Apparatus and Procedure}
The cell (described in  more detail in ref.~\onlinecite{Ray2009b})  used for this experiment is shown schematically in figure \ref{fig:cell}.  It consisted of a cylindrical copper chamber (V = 1.8 cm$^3$), where the solid was grown, pierced by two Vycor rods.  It should be noted that the data in this report come from two different sets of Vycor rods.  The first set was made from rods that were 1.5 mm in diameter, and was the same Vycor as was used in references \onlinecite{Ray2008a,Ray2009,Ray2009b,,Ray2010}.  Additional data was also taken using Vycor that was 3.0 mm in diameter.  In the latter, the Vycor was not contained in stainless steel tubes as before, but rather, was encased in Stycast 2850 FT epoxy.  The change to epoxy encasement was done to ensure that there could be no possibility of parallel pathways for helium to bypass the Vycor.  Two capacitance strain gages\cite{Straty1969}, C1 and C2, were affixed to the ends of the cell to measure the pressure of the solid.  These two pressure gauges allowed us to independently measure the pressure in the solid at each end of the cell and thus identify any pressure gradients that might exist across the solid.  The cell was bolted onto a copper plate that was attached to the mixing chamber of a dilution refrigerator by means of solid copper bars.

To initiate the growth of solid helium from the superfluid at constant temperature, we simultaneously admit helium to the cell initially at $P < 25$ bar via lines 1 and 2 and the pressure in the cell increases until it reaches the freezing pressure, $P_{F}$, at which liquid and solid coexist.  At the melting pressure, $dP/dt = 0$ since $P = P_F$ as long as there is solid and liquid in coexistence (and no solid regions form that bridge the cell diameter).  During solid growth and subsequent addition of atoms to the cell, $P_1, P_2$ where typically in the range between $26.5$ and $27.2$ bar, which results in the addition of atoms to the cell at a typical rate of $dN/dt \approx 2 \times 10^{16}$ atoms/s.  As we continue to add atoms to the cell through the Vycor rods, we observe the cell pressure to rise above $P_F$.  The continued addition of atoms to the cell increases the density of the solid.

\begin{figure}
\resizebox{3 in}{!}{
\includegraphics{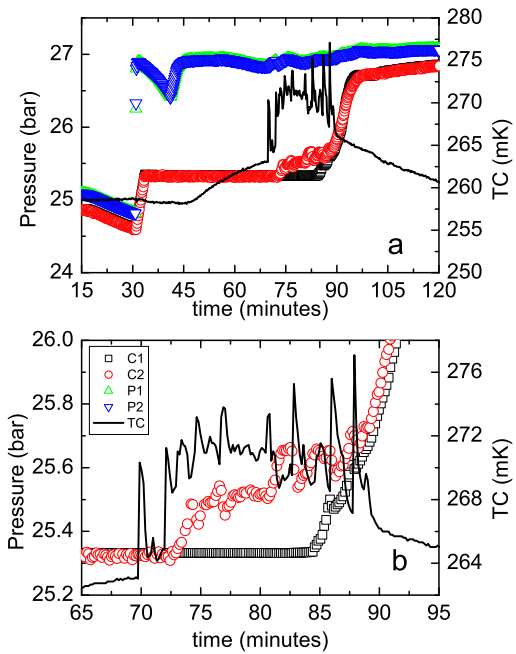}}
\caption{\label{fig:FI}  (color online) Growth of solid helium from superfluid at $TC \approx 260$ mK; sample FI.  (a) Complete growth record showing all four pressures and TC.  (b) Closeup of the transients seen on and near the melting curve.  Most of the transient increases in temperature are accompanied by drops in the pressure measured by C1 and C2. }
\end{figure}
\subsection{Transient Events}
Figure \ref{fig:FI} shows one such growth record, the growth of solid sample FI. Notice first in figure \ref{fig:FI}a, that when both $C1 = C2 = P_F$ ($t <$ 65 min) the temperature recorded at TC is smooth, then shortly before C2 comes off the melting curve the temperature starts to fluctuate by $\sim 10$mK.  These fluctuations persist until shortly after C1 comes off the melting curve at which point the signal becomes smooth again. This change in behavior of TC is perhaps due to the way in which the solid grows.  Initially, before the solid has bridged or filled the cell, it can grow uniformly from the liquid-solid free surface, with liquid regions connected throughout the cell.  The increase of the pressure above the bulk melting curve indicates the cell has filled with solid, or that regions of the cell are separated by bulk solid and a connected solid-liquid surface no longer exists. Once the cell is entirely filled with solid, the solid must find a new method to grow - by increasing its density.  As the density of the solid increases, the probability that any liquid inclusions imbedded in the solid will solidify increases.  Such meta-stable liquid regions have been previously observed in solid helium \cite{Mikhin2001,Mikhin2007,Pantalei2010}.  When one of these regions does solidify, energy is released, the local density rises and the measured pressure falls.  We believe that this is what causes the transients, which can be seen in greater detail in figure \ref{fig:FI}b.  When the pressure drops cease, the temperature record becomes smooth.

\begin{figure}
\resizebox{3 in}{!}{
\includegraphics{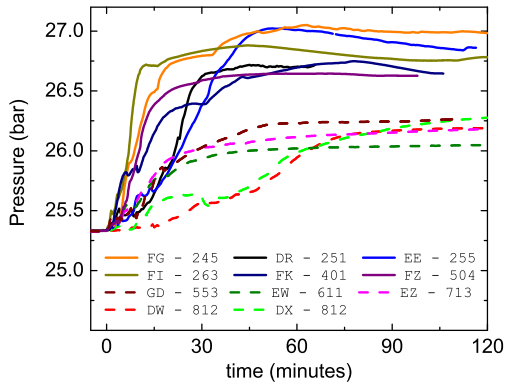}}
\caption{\label{fig:growth-comp} (color online) Dependence of C1 on time for the growth of several different samples at various temperatures (indicated in mK in the legend).  Data from samples at temperatures above 550 mK are shown with dashed lines.  The fill line pressures during the growth varied from 26.51 to 27.20 bar. The time axis has been shifted for each data set so that the pressure first rises above the bulk melting curve at t = 0.}
\end{figure}

We observe these transients when the pressure of the solid is within $\sim 0.5$ bar of the melting curve, and at temperatures $TC \leq 550$ mK.  At temperatures higher than this, we sometimes observe drops in the pressure, but they are not accompanied by a resolvable associated temperature transient.  Furthermore, as shown in figure \ref{fig:growth-comp} (which shows data taken with the larger diameter Vycor described above), there is a difference in behavior between samples grown at $TC \approx 550$ mK and above, and those grown at lower temperatures.  Below $\approx 550$ mK when the solid departs the bulk melting curve the pressure measured by C1 and C2 rises quickly ($\sim 15$ to $60$ minutes) to near the pressure of the fill lines.  Above 500 mK, the pressure in the cell seems to level off between 26.10 and 26.25 bar, regardless of the pressure in the fill lines.  In other words, above $TC \approx 500$ mK, we can only grow solid samples to $\sim 26.25$ bar.  As reported previously \cite{Ray2008a,Ray2009,Ray2009b,Ray2010}, at these higher temperatures, we also observe no mass flow.



We first assume that these transient events are in fact due to the solidification of over-pressurized liquid regions and calculate the size of the region that solidified considering the idealized case of a single such event. Initially,
before the pressure drop, we have some mass of solid, $m_S$, and some mass of liquid $m_L$ so that the total mass contained in
region S, $m_i$, is $m_i = m_S + m_L = \rho_s V_S + \rho_L V_L$,
where $\rho_S$ and $\rho_L$, and V$_S$ and V$_L$ are the density
and volume of the solid and liquid, respectively.  If, during the
transient, all the liquid, $m_L$, was converted to solid, then the final
volume occupied by the solid is equal to the cell volume,
V$_{cell}$, so that $m_f = \rho'_S V_{cell}$. Where $\rho'_S$ is
the density of the solid after the pressure drop. This
makes the assumption that the density is uniform throughout the
solid. Immediately after the transient, but before extra mass from
the Vycor has entered the cell, the total mass in the cell has not
changed so that $m_f$ = $m_i$.  Solving for V$_L$
 (keeping in mind that the initial volume of solid is $V_S =
V_{cell} - V_L$), we find
\begin{equation}
\label{eq:dV}
V_L \approx \frac{(\rho_S - \rho'_S)}{(\rho_S - \rho_L)}V_{cell}.
\end{equation}

We can take an such event and use it to estimate the volume of liquid that solidified during the event.  Take for example the transient in figure \ref{fig:FI} at t = 88 min. The pressure measured on C1 immediately before the transient is $P_i = 25.714$ bar corresponding to a density of $\rho_S = 0.1899$ g/cm$^3$. The pressure then dropped to $P_f = 25.662$ bar corresponding to a final density of $\rho'_S = 0.1901$ g/cm$^3$.  Using $\rho_L = 0.1724$ g/cm$^3$, which is an extrapolation of the density of liquid helium to $P_i$, we can deduce that a volume $dV \approx 3.6 \times 10^{-3}$ cm$^3$ converted from liquid to solid. This represents about $0.2 \%$ of the volume of region S.

\begin{figure}
\resizebox{3 in}{!}{
\includegraphics{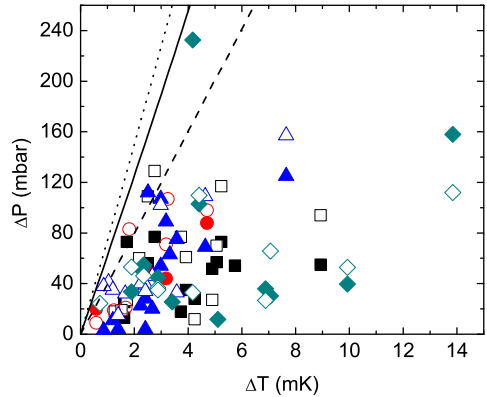}}
\caption{\label{fig:DPDT} (Color online) Pressure drop, $\Delta P$
versus temperature rise, $\Delta T$ for transients at several
temperatures.  Filled symbols refer to C1; open refer to C2.
Squares, 260 mK; circles, 370 mK; triangles, 400 mK; diamonds, 500 mK. Lines represent the expected relationships (see text). Solid line: $P_L = P_S$ (no strain field); Dashed line: $P_L = P_F$, $R_{strain} = 3 r_L$; Dotted line: $P_L = P_F$, $R_{strain} = 2 r_L$.}
\end{figure}

One should also be able to use the temperature change to calculate the amount of liquid that solidified.  A logical start is to assume that the temperature transient associated
with the pressure drop is due to the latent heat released
upon solidification of the liquid region.  However, because the
latent heat of solidification for helium is so small
($~10^{-4}$J/mol \cite{Swenson1950}), one finds that liquid helium in almost half
of the volume of the cell would have had to have solidified at once in order to
account for the temperature rise.  This is quite unlikely since we
typically see multiple transients during the growth of a single sample.

Alternatively, the temperature transient associated with the
pressure drop could be due to the work done in expanding the
volume of the solid by an amount $dV$. If we denote the thermal energy
associated with an event as $dQ$, we can write $dQ = d(E +PV)$,
where $E$ is the internal energy and $P$ and $V$ are the pressure
and volume of the solid before the event.  Assuming the internal
energy does not change significantly, and writing $dQ = dT[\Sigma
M_{i}c_{i}]$, where $M_i$ is the mass of each component whose
temperature is taken to rise by $dT$ (the helium, the cell and the
mixing chamber), and $c_i$ is the heat capacity of each element,
we have an expression relating the temperature rise due to a pressure drop:
\begin{equation}
\label{eq:dT}
dT \approx (PdV + VdP)/(\Sigma M_{i}c_{i}),
\end{equation}
where $dV$ can be calculated for a given pressure drop using eq.~\ref{eq:dV}.

Figure \ref{fig:DPDT} shows the observed pressure drop measured in C1 and C2, $\Delta P$, versus the temperature rise recorded on TC,
$\Delta T$, for several temperatures along with eq.~\ref{eq:dT} (solid line), for the case of meta-stable
over-pressurized liquid. There is a spread in the data, but in
general the pressure drop does in fact increase with the observed
temperature rise.  The spread
in the data likely means that the events are localized within the
solid at various distances from the capacitors, and so the full
pressure drop is not resolved on the capacitors; thus, the measured $\Delta P$ is typically
smaller than the size predicted by eq. 2. The local nature of each event is
further demonstrated by the observation that the two capacitors
often register different pressure drops for the same event.

Until this point we have assumed that the pressure of the liquid region is equal to that of the solid.  However, we could also adopt the view that the meta-stable liquid regions are at a lower pressure. In that case a strained region in the solid will exist and we should include the energy involved in the change of the density of the strained solid around the liquid region when the liquid solidifies. In the most extreme case the liquid might be at the melting curve pressure. On figure 4 we show one case (dotted line) where the strained solid is assumed to extend to two times the radius of the liquid region, $r_L$, that has volume $dV$ calculated from eq.~\ref{eq:dV}, and one case (dashed line) where the strained region extends to  3$r_L$ (this is the maximum spherical size that will fit in the cell). With this, it appears that the solidification of liquid regions is certainly involved, but given the uncertainty of the extent of the stain field and the density, $\rho_L$, of the liquid region, the relationship can't be
precisely deduced.

\subsection{Post-transient Growth: Syringe Experiments}
Next we focus on the further growth of solid helium when $P > P_F$. First,we note that two types of measurements are possible with the apparatus: (1) inject atoms into either line 1 or 2 and monitor
the other for evidence that mass has moved through the solid helium, while also observing C1 and C2, or (2) inject via
lines 1 and 2 simultaneously and simply observe the behavior of C1 and C2.
We have reported on experiments of the first type\cite{Ray2008a,
Ray2009, Ray2009b,Ray2010}, where changes in C1 and C2 are only observed
when we also see evidence for the flow of mass through the solid.
The second type of experiment, which has been termed
a superfluid syringe\cite{Soyler2009}, is useful to further study how the solid can grow isochorically, i.e. how the solid can increase its density at fixed cell volume.

\begin{figure}
\resizebox{3 in}{!}{
\includegraphics{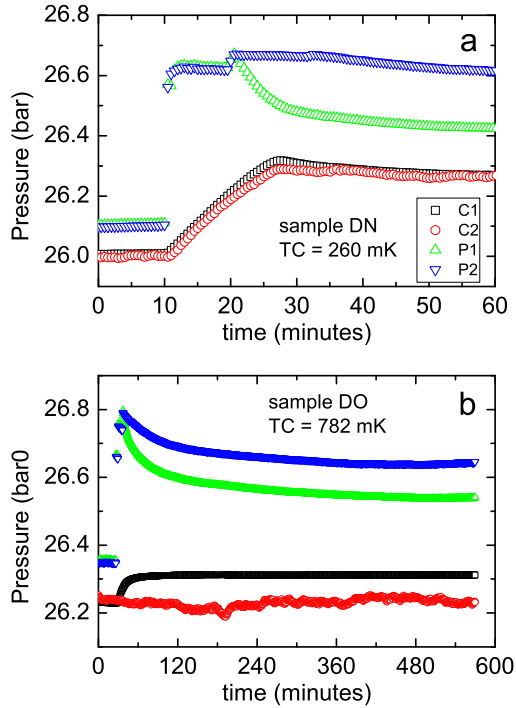}}
\caption{\label{fig:DN-DO}(color online) (a) A syringe type experiment that shows a large isochoric compressibility.  The pressure in both reservoirs is increased at the same time and a corresponding rise in the pressure measured by C1 and C2 is recorded.  (b) The sample was then warmed to 782 mK, and a second syringe-type push showed no long term response in either C1 or C2.}
\end{figure}

Two examples of the syringe experiment are shown in figure
\ref{fig:DN-DO}.  Figure \ref{fig:DN-DO}a portrays data from sample DN, grown fresh from the superfluid at 260 mK.  Following growth, P1 and P2 were increased simultaneously by $0.520$ bar, injecting atoms into the solid.  We continued to inject atoms for 10 minutes, then shut off the regulator, and closed lines 1 and 2.  C1 and C2 both registered corresponding increases in the pressure of the solid.  Note that after the valves feeding atoms to the Vycor were closed, the pressure in line 1 decreased to equilibrium with the cell pressure within $\sim 10$ minutes, but line 2 fell much more slowly.  This could indicate a difference in the flow through the two Vycor rods, but regardless, the two capacitors rose at the same rate, meaning that even if there was a flow rate difference through the Vycor, the solid was conducting atoms.

Sample DN was then warmed up to 782 mK to create sample DO and injected in the same way as DN with the pressure raised by $0.312$ bar.  After a short-term rise in C1 (shown in \ref{fig:DN-DO}b), there was no long-term evidence of mass entering the solid.  There was also no response in C2, indicating that there was no mass movement across the cell.  With the 3 mm diameter Vycor, we observe this short term behavior in C1 (e.g. near $t \approx$ 40 min) when we increase P1 at temperatures above 600 mK.  This behavior is different from that we have reported in our previous experiments \cite{Ray2009b} where we saw no response in either capacitor at $TC \gtrsim 600$ mK.  Interestingly, this response is mostly seen in C1, when P1 is increased, though a slower response can sometimes be seen in C2 over a long period of time.  If only P2 is increased at these higher temperatures, then very little response in either capacitor can be observed. In either case there is no flow between the Vycor ``electrodes."

The nature of the behavior of C1, and why it is somewhat different from C2 is not entirely clear to us.  There is the possibility that the mass flux through V2 is smaller than V1, and C1 responds more quickly because it is closer to the Vycor that is conducting more mass.  However, we see no such asymmetry at lower temperatures, with the same Vycor reservoir temperatures.   In other words, a sample created at $TC \approx 250$ mK, and with Vycor temperatures $TV1$ and $TV2$, show the same behavior when increasing the pressure in each reservoir separately.  But when the cell is warmed to $TC \approx 700$ mK, with $TV1$ and $TV2$ unchanged, we observe short term behavior in C1, but not in C2 when only P1 is increased, and no short term behavior in either C1 or C2 when only P2 is increased.  This makes it unlikely that the asymmetry is due to the Vycor itself.  We doubt that this short term behavior is due to the frost-heave effect which has been observed \cite{Hiroi1989} because frost heave is not consistent with the observed asymmetry and such behavior was not seen in our earlier work\cite{Ray2009b}.

Ignoring the short term changes in C1, and the asymmetry in the solid's behavior, it is clear the response of the solid to an increase in P1 and P2 at $TC = 250$ mK is very different than at $TC \geq 600$ mK.  At the lower temperatures both capacitors rise quickly in response to the pressure increase, and there is a relatively sharp cutoff in $dC/dt$ when equilibrium between the cell and reservoirs is achieved.  At the higher temperatures, the response is slower, and much more rounded.  It is also seen that at $TC \geq 600$ mK there is no mass flux across the cell, which is consistent with our previous observations that the solid does not support a mass flow at these higher temperatures.

\begin{figure}
\resizebox{3 in}{!}{
\includegraphics{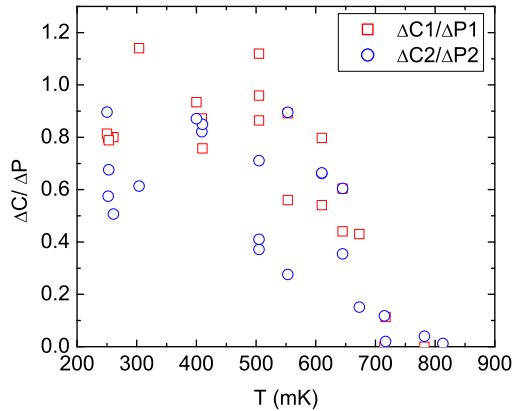}}
\caption{\label{fig:DP} (color online) Change in cell pressure divided by the the change in reservoir (fill line) pressure.  The data have been adjusted to account for the short term behavior seen in C1.}
\end{figure}

Soyler {\it et al.}~\cite{Soyler2009} have developed a theory of how solid helium can grow at $P > P_F$, which could explain the relationship between our observance of flow and changes in the cell pressure \cite{Ray2009b}.  The basic idea presented in the theory is that the solid can only grow by the climb of edge dislocations where mass is fed to the dislocation along superfluid cores~\cite{Soyler2009}.  When there is no flow along the dislocations, the solid cannot grow, and must be incompressible.  This model requires that the temperature be low enough so that the dislocation cores are superfluid, but high enough so that the dislocations are rough, which means there is a finite temperature range in which one should observe isochoric compressibility of the solid.

In ref.~\onlinecite{Soyler2009} the isochoric compressibility is defined as $\chi = dn/d\mu$, where $dn$ is the density change in the solid, in response to the chemical potential change in the fill lines, $d\mu$, when the pressure is increased.  Figure \ref{fig:DP} shows the analog of $\chi$, $\Delta C1 / \Delta P1$ and $\Delta C2 / \Delta P2$, a measure of the compressibility of the solid for non-thermally cycled samples over a range of temperatures and pressures.  (Thermal cycles can change the flow behavior\cite{Ray2009b}.)  The data have been adjusted so that when $\Delta C2 / \Delta P2 < 0.05$ we set $\Delta C1 / \Delta P1 = 0$ to account for the short term behavior in C1 (Figure 5b, $t \approx$ 40 min.). As can be seen, $\Delta C/ \Delta P$ appears to rise with an increase in temperature, have a plateau or maximum near $T$ = 400 mK, decrease strongly with an increase in temperature above 500 mK, and approach 0 at $TC \approx 700$ mK.  This qualitative behavior is fully consistent with the predictions of Soyler {\it et al}~\cite{Soyler2009}.  However, to confirm this theory unambiguously requires a reconstruction of our apparatus and an extension of our results to lower temperatures to search for the predicted lower cutoff temperature below which the solid is predicted to continue to demonstrate flow as a result of an applied chemical potential difference, but no increase in density.  We currently have limited data that shows increases of density and flow for $TC \approx 120$ mK, so any lower cutoff temperature must be below this.

\begin{figure}
\resizebox{3 in}{!}{
\includegraphics{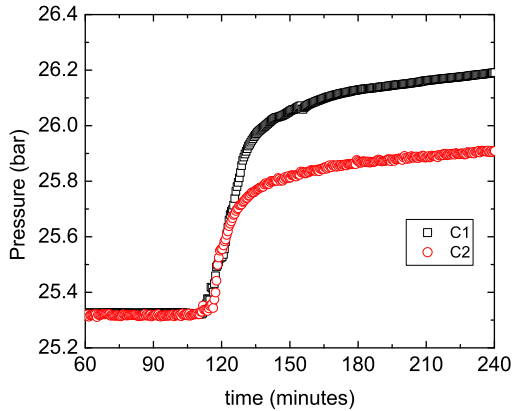}}
\caption{\label{fig:EZ} (color online) Growth of sample EZ from the superfluid at $TC = 712$ mK.  In this sample, a long-term pressure gradient appeared between the two pressure gages.}
\end{figure}
\subsection{Pressure Gradients}
Finally, we comment on the pressure measured by the capacitors on either end of the cell.  Note that in figure \ref{fig:FI}, $C1 = C2$ throughout most of the data record.  However, this is not always the case, and we often see pressure differences appear between the two capacitors\cite{Ray2009b}.  Figure \ref{fig:EZ} shows the growth of a solid sample (EZ) at $TC = 712$ mK in which a pressure difference appeared in the solid.  These pressure differences can be quite large, sometimes reaching as high as $\sim 200$ mbar.  Also, in figure \ref{fig:FI}, C2 starts to measure pressures higher than $P_F$ 13 minutes before C1 registers pressures greater than $P_F$ meaning that even with liquid in the solid chamber pressure gradients can occur across the cell, presumably due to the isolation of liquid regions by the solid.  Although it is tempting to think of this pressure difference as a gradient, this gives the impression of smoothly varying pressure between the two capacitors.  We suspect that there is likely fluctuation in the pressure throughout the solid; there is no need for the pressure to vary smoothly since defects in the solid can allow the solid to have local pressure gradients.

\begin{figure}
\resizebox{3 in}{!}{
\includegraphics{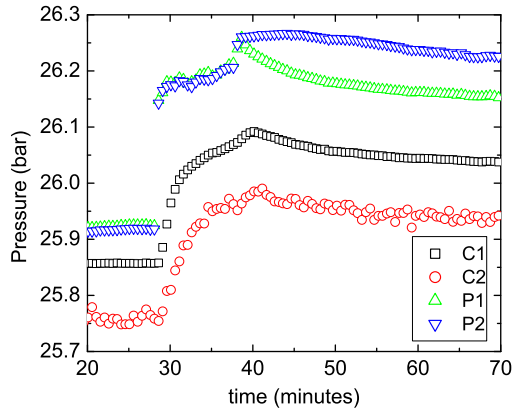}}
\caption{\label{fig:DR} (color online) Sample DR, T = 253 mK - a syringe-type experiment done on a solid sample with a stable pressure gradient across the solid of $\approx 150$ mbar.}
\end{figure}

Although performing a flow or syringe experiment can sometimes alter the size of the pressure difference by several percent, it usually persists throughout the life of the solid sample ($\sim 2$ days).  Further, due to the asymmetry in the behavior across the solid, a syringe experiment done at temperatures where there is no mass flow will create a pressure difference, as is the case in figure \ref{fig:DN-DO}b.  As reported before, the existence of a pressure difference across the solid does not affect whether or not flow is observed \cite{Ray2009b}.  To demonstrate this, figure \ref{fig:DR} shows a syringe-type experiment done on a sample with a pressure difference of $\Delta C \approx 100$ mbar at $TC = 253$ mK (sample DR).  Compared to figure \ref{fig:DN-DO}a where there was no pressure difference it is apparent that the pressure gradients do not effect the properties of the flow response of the solid.

\section{Conclusion}

In conclusion, we have studied the growth of solid helium at $P > P_F$ by injecting atoms into the solid.  When injecting atoms near the melting curve, we observed transients in the cell temperature accompanied by pressure drops.  The pressure drops vary in size up to $\sim 100$ mbar, and the temperature transients are are up to $\sim 10$ mK.  These events are not seen in samples grown at higher temperature ($TC \gtrsim 550$ mK). We believe that
these events are due to liquid regions trapped within the
solid that solidify. Further from the melting curve such
transients are not seen; there the addition of atoms to the cell
results in the growth of the density of the solid at constant
volume, this growth perhaps may be understood in terms of the superclimb of edge dislocations\cite{Soyler2009}.  Finally, we often observe pressure differences are present in solid $^4$He between our two pressure gauges located on the edges of the cell.  These pressure difference occur during the growth of the solid sample, and persist throughout the life of the sample.  These pressure differences have no effect on the response of the solid to mass injections. The fact that we observe this pressure difference between the two capacitors means that the pressure (and density) is probably quite inhomogeneous throughout the solid.  Of course, this suggests that experiments that utilize a single pressure gauge to study samples that are not annealed may not be able to accurately locate such samples on the phase diagram.

\section{Acknowledgements}

We thank S. Balibar, J. Beamish, D. Huse, N. Prokofev and B.
Svistunov for discussions. This work was supported by NSF DMR
06-50096 and 08-55954 and to a limited extent by 07-57701.

\bibliography{ref}

\begin{thebibliography}{29}%
\makeatletter
\providecommand \@ifxundefined [1]{%
 \@ifx{#1\undefined}
}%
\providecommand \@ifnum [1]{%
 \ifnum #1\expandafter \@firstoftwo
 \else \expandafter \@secondoftwo
 \fi
}%
\providecommand \@ifx [1]{%
 \ifx #1\expandafter \@firstoftwo
 \else \expandafter \@secondoftwo
 \fi
}%
\providecommand \natexlab [1]{#1}%
\providecommand \enquote  [1]{``#1''}%
\providecommand \bibnamefont  [1]{#1}%
\providecommand \bibfnamefont [1]{#1}%
\providecommand \citenamefont [1]{#1}%
\providecommand \href@noop [0]{\@secondoftwo}%
\providecommand \href [0]{\begingroup \@sanitize@url \@href}%
\providecommand \@href[1]{\@@startlink{#1}\@@href}%
\providecommand \@@href[1]{\endgroup#1\@@endlink}%
\providecommand \@sanitize@url [0]{\catcode `\\12\catcode `\$12\catcode
  `\&12\catcode `\#12\catcode `\^12\catcode `\_12\catcode `\%12\relax}%
\providecommand \@@startlink[1]{}%
\providecommand \@@endlink[0]{}%
\providecommand \url  [0]{\begingroup\@sanitize@url \@url }%
\providecommand \@url [1]{\endgroup\@href {#1}{\urlprefix }}%
\providecommand \urlprefix  [0]{URL }%
\providecommand \Eprint [0]{\href }%
\@ifxundefined \urlstyle {%
  \providecommand \doi  [0]{\begingroup \@sanitize@url \@doi}%
  \providecommand \@doi [1]{\endgroup \@@startlink {\doibase
  #1}doi:\discretionary {}{}{}#1\@@endlink }%
}{%
  \providecommand \doi  [0]{doi:\discretionary{}{}{}\begingroup
  \urlstyle{rm}\Url }%
}%
\providecommand \doibase [0]{http://dx.doi.org/}%
\providecommand \Doi [0]{\begingroup \@sanitize@url \@Doi }%
\providecommand \@Doi  [1]{\endgroup\@@startlink{\doibase#1}\@@Doi}%
\providecommand \@@Doi [1]{#1\@@endlink}%
\providecommand \selectlanguage [0]{\@gobble}%
\providecommand \bibinfo  [0]{\@secondoftwo}%
\providecommand \bibfield  [0]{\@secondoftwo}%
\providecommand \translation [1]{[#1]}%
\providecommand \BibitemOpen [0]{}%
\providecommand \bibitemStop [0]{}%
\providecommand \bibitemNoStop [0]{.\EOS\space}%
\providecommand \EOS [0]{\spacefactor3000\relax}%
\providecommand \BibitemShut  [1]{\csname bibitem#1\endcsname}%
\bibitem [{\citenamefont {Andreev}\ and\ \citenamefont
  {Lifshitz}(1969)}]{Andreev1969}%
  \BibitemOpen
  \bibfield  {author} {\bibinfo {author} {\bibfnamefont {A.}~\bibnamefont
  {Andreev}}\ and\ \bibinfo {author} {\bibfnamefont {I.}~\bibnamefont
  {Lifshitz}},\ }\href@noop {} {\bibfield  {journal} {\bibinfo  {journal} {Sov.
  Phys. JETP},\ }\textbf {\bibinfo {volume} {29}},\ \bibinfo {pages} {1107}
  (\bibinfo {year} {1969})}\BibitemShut {NoStop}%
\bibitem [{\citenamefont {Chester}(1970)}]{Chester1970}%
  \BibitemOpen
  \bibfield  {author} {\bibinfo {author} {\bibfnamefont {G.~V.}\ \bibnamefont
  {Chester}},\ }\Doi {10.1103/PhysRevA.2.256} {\bibfield  {journal} {\bibinfo
  {journal} {Phys. Rev. A},\ }\textbf {\bibinfo {volume} {2}},\ \bibinfo
  {pages} {256} (\bibinfo {year} {1970})}\BibitemShut {NoStop}%
\bibitem [{\citenamefont {Leggett}(1970)}]{Leggett1970}%
  \BibitemOpen
  \bibfield  {author} {\bibinfo {author} {\bibfnamefont {A.~J.}\ \bibnamefont
  {Leggett}},\ }\Doi {10.1103/PhysRevLett.25.1543} {\bibfield  {journal}
  {\bibinfo  {journal} {Phys. Rev. Lett.},\ }\textbf {\bibinfo {volume} {25}},\
  \bibinfo {pages} {1543} (\bibinfo {year} {1970})}\BibitemShut {NoStop}%
\bibitem [{\citenamefont {Ho}\ \emph {et~al.}(1997)\citenamefont {Ho},
  \citenamefont {Bindloss},\ and\ \citenamefont {Goodkind}}]{Ho1997}%
  \BibitemOpen
  \bibfield  {author} {\bibinfo {author} {\bibfnamefont {P.}~\bibnamefont
  {Ho}}, \bibinfo {author} {\bibfnamefont {I.}~\bibnamefont {Bindloss}}, \ and\
  \bibinfo {author} {\bibfnamefont {J.}~\bibnamefont {Goodkind}},\ }\Doi
  {10.1007/s10909-005-0094-0} {\bibfield  {journal} {\bibinfo  {journal} {J.
  Low. Temp. Phys.},\ }\textbf {\bibinfo {volume} {109}},\ \bibinfo {pages}
  {409} (\bibinfo {year} {1997})}\BibitemShut {NoStop}%
\bibitem [{\citenamefont {Kim}\ and\ \citenamefont
  {Chan}(2004){\natexlab{a}}}]{Kim2004a}%
  \BibitemOpen
  \bibfield  {author} {\bibinfo {author} {\bibfnamefont {E.}~\bibnamefont
  {Kim}}\ and\ \bibinfo {author} {\bibfnamefont {M.}~\bibnamefont {Chan}},\
  }\href@noop {} {\bibfield  {journal} {\bibinfo  {journal} {Nature},\ }\textbf
  {\bibinfo {volume} {427}},\ \bibinfo {pages} {225} (\bibinfo {year}
  {2004}{\natexlab{a}})}\BibitemShut {NoStop}%
\bibitem [{\citenamefont {Kim}\ and\ \citenamefont
  {Chan}(2004){\natexlab{b}}}]{Kim2004b}%
  \BibitemOpen
  \bibfield  {author} {\bibinfo {author} {\bibfnamefont {E.}~\bibnamefont
  {Kim}}\ and\ \bibinfo {author} {\bibfnamefont {M.}~\bibnamefont {Chan}},\
  }\href@noop {} {\bibfield  {journal} {\bibinfo  {journal} {Science},\
  }\textbf {\bibinfo {volume} {305}},\ \bibinfo {pages} {1941} (\bibinfo {year}
  {2004}{\natexlab{b}})}\BibitemShut {NoStop}%
\bibitem [{\citenamefont {Kim}\ and\ \citenamefont {Chan}(2005)}]{Kim2005}%
  \BibitemOpen
  \bibfield  {author} {\bibinfo {author} {\bibfnamefont {E.}~\bibnamefont
  {Kim}}\ and\ \bibinfo {author} {\bibfnamefont {M.}~\bibnamefont {Chan}},\
  }\href@noop {} {\bibfield  {journal} {\bibinfo  {journal} {J. Low Temp.
  Phys.},\ }\textbf {\bibinfo {volume} {138}},\ \bibinfo {pages} {859}
  (\bibinfo {year} {2005})}\BibitemShut {NoStop}%
\bibitem [{\citenamefont {Balibar}\ and\ \citenamefont
  {Caupin}(2008)}]{Balibar2008}%
  \BibitemOpen
  \bibfield  {author} {\bibinfo {author} {\bibfnamefont {S.}~\bibnamefont
  {Balibar}}\ and\ \bibinfo {author} {\bibfnamefont {F.}~\bibnamefont
  {Caupin}},\ }\href@noop {} {\bibfield  {journal} {\bibinfo  {journal} {J.
  Phys.: Condens. Matter},\ }\textbf {\bibinfo {volume} {20}},\ \bibinfo
  {pages} {173201} (\bibinfo {year} {2008})}\BibitemShut {NoStop}%
\bibitem [{\citenamefont {Day}\ and\ \citenamefont {Beamish}(2007)}]{Day2007}%
  \BibitemOpen
  \bibfield  {author} {\bibinfo {author} {\bibfnamefont {J.}~\bibnamefont
  {Day}}\ and\ \bibinfo {author} {\bibfnamefont {J.}~\bibnamefont {Beamish}},\
  }\Doi {10.1038/nature06383} {\bibfield  {journal} {\bibinfo  {journal}
  {Nature},\ }\textbf {\bibinfo {volume} {450}},\ \bibinfo {pages} {853}
  (\bibinfo {year} {2007})}\BibitemShut {NoStop}%
\bibitem [{\citenamefont {Day}\ \emph {et~al.}(2010)\citenamefont {Day},
  \citenamefont {Syshchenko},\ and\ \citenamefont {Beamish}}]{Day2010}%
  \BibitemOpen
  \bibfield  {author} {\bibinfo {author} {\bibfnamefont {J.}~\bibnamefont
  {Day}}, \bibinfo {author} {\bibfnamefont {O.}~\bibnamefont {Syshchenko}}, \
  and\ \bibinfo {author} {\bibfnamefont {J.}~\bibnamefont {Beamish}},\ }\Doi
  {10.1103/PhysRevLett.104.075302} {\bibfield  {journal} {\bibinfo  {journal}
  {Phys. Rev. Lett.},\ }\textbf {\bibinfo {volume} {104}},\ \bibinfo {pages}
  {075302} (\bibinfo {year} {2010})}\BibitemShut {NoStop}%
\bibitem [{\citenamefont {Rojas}\ \emph {et~al.}(2010)\citenamefont {Rojas},
  \citenamefont {Pantalei}, \citenamefont {Maris},\ and\ \citenamefont
  {Balibar}}]{Rojas2010}%
  \BibitemOpen
  \bibfield  {author} {\bibinfo {author} {\bibfnamefont {X.}~\bibnamefont
  {Rojas}}, \bibinfo {author} {\bibfnamefont {C.}~\bibnamefont {Pantalei}},
  \bibinfo {author} {\bibfnamefont {H.}~\bibnamefont {Maris}}, \ and\ \bibinfo
  {author} {\bibfnamefont {S.}~\bibnamefont {Balibar}},\ }\Doi
  {10.1007/s10909-009-9966-z} {\bibfield  {journal} {\bibinfo  {journal} {J.
  Low Temp. Phys.},\ }\textbf {\bibinfo {volume} {158}},\ \bibinfo {pages}
  {478} (\bibinfo {year} {2010})}\BibitemShut {NoStop}%
\bibitem [{\citenamefont {Greywall}(1977)}]{Greywall1977}%
  \BibitemOpen
  \bibfield  {author} {\bibinfo {author} {\bibfnamefont {D.~S.}\ \bibnamefont
  {Greywall}},\ }\Doi {10.1103/PhysRevB.16.1291} {\bibfield  {journal}
  {\bibinfo  {journal} {Phys. Rev. B},\ }\textbf {\bibinfo {volume} {16}},\
  \bibinfo {pages} {1291} (\bibinfo {year} {1977})}\BibitemShut {NoStop}%
\bibitem [{\citenamefont {Day}\ \emph {et~al.}(2005)\citenamefont {Day},
  \citenamefont {Herman},\ and\ \citenamefont {Beamish}}]{Day2005}%
  \BibitemOpen
  \bibfield  {author} {\bibinfo {author} {\bibfnamefont {J.}~\bibnamefont
  {Day}}, \bibinfo {author} {\bibfnamefont {T.}~\bibnamefont {Herman}}, \ and\
  \bibinfo {author} {\bibfnamefont {J.}~\bibnamefont {Beamish}},\ }\Doi
  {10.1103/PhysRevLett.95.035301} {\bibfield  {journal} {\bibinfo  {journal}
  {Phys. Rev. Lett.},\ }\textbf {\bibinfo {volume} {95}},\ \bibinfo {eid}
  {035301} (\bibinfo {year} {2005})}\BibitemShut {NoStop}%
\bibitem [{\citenamefont {Day}\ and\ \citenamefont {Beamish}(2006)}]{Day2006}%
  \BibitemOpen
  \bibfield  {author} {\bibinfo {author} {\bibfnamefont {J.}~\bibnamefont
  {Day}}\ and\ \bibinfo {author} {\bibfnamefont {J.}~\bibnamefont {Beamish}},\
  }\Doi {10.1103/PhysRevLett.96.105304} {\bibfield  {journal} {\bibinfo
  {journal} {Phys. Rev. Lett.},\ }\textbf {\bibinfo {volume} {96}},\ \bibinfo
  {eid} {105304} (\bibinfo {year} {2006})}\BibitemShut {NoStop}%
\bibitem [{\citenamefont {Rittner}\ \emph {et~al.}(2009)\citenamefont
  {Rittner}, \citenamefont {Choi}, \citenamefont {Mueller},\ and\ \citenamefont
  {Reppy}}]{Rittner2009}%
  \BibitemOpen
  \bibfield  {author} {\bibinfo {author} {\bibfnamefont {A.~S.~C.}\
  \bibnamefont {Rittner}}, \bibinfo {author} {\bibfnamefont {W.}~\bibnamefont
  {Choi}}, \bibinfo {author} {\bibfnamefont {E.~J.}\ \bibnamefont {Mueller}}, \
  and\ \bibinfo {author} {\bibfnamefont {J.~D.}\ \bibnamefont {Reppy}},\ }\Doi
  {10.1103/PhysRevB.80.224516} {\bibfield  {journal} {\bibinfo  {journal}
  {Phys. Rev. B},\ }\textbf {\bibinfo {volume} {80}},\ \bibinfo {pages}
  {224516} (\bibinfo {year} {2009})}\BibitemShut {NoStop}%
\bibitem [{\citenamefont {Ray}\ and\ \citenamefont {Hallock}(2008)}]{Ray2008a}%
  \BibitemOpen
  \bibfield  {author} {\bibinfo {author} {\bibfnamefont {M.~W.}\ \bibnamefont
  {Ray}}\ and\ \bibinfo {author} {\bibfnamefont {R.~B.}\ \bibnamefont
  {Hallock}},\ }\Doi {10.1103/PhysRevLett.100.235301} {\bibfield  {journal}
  {\bibinfo  {journal} {Phys. Rev. Lett.},\ }\textbf {\bibinfo {volume}
  {100}},\ \bibinfo {eid} {235301} (\bibinfo {year} {2008})}\BibitemShut
  {NoStop}%
\bibitem [{\citenamefont {Ray}\ and\ \citenamefont
  {Hallock}(2009){\natexlab{a}}}]{Ray2009}%
  \BibitemOpen
  \bibfield  {author} {\bibinfo {author} {\bibfnamefont {M.~W.}\ \bibnamefont
  {Ray}}\ and\ \bibinfo {author} {\bibfnamefont {R.~B.}\ \bibnamefont
  {Hallock}},\ }\Doi {10.1088/1742-6596/150/3/032087} {\bibfield  {journal}
  {\bibinfo  {journal} {J. Phys.: Conf. Ser.},\ }\textbf {\bibinfo {volume}
  {150}},\ \bibinfo {pages} {032087} (\bibinfo {year}
  {2009}{\natexlab{a}})}\BibitemShut {NoStop}%
\bibitem [{\citenamefont {Ray}\ and\ \citenamefont
  {Hallock}(2009){\natexlab{b}}}]{Ray2009b}%
  \BibitemOpen
  \bibfield  {author} {\bibinfo {author} {\bibfnamefont {M.~W.}\ \bibnamefont
  {Ray}}\ and\ \bibinfo {author} {\bibfnamefont {R.~B.}\ \bibnamefont
  {Hallock}},\ }\Doi {10.1103/PhysRevB.79.224302} {\bibfield  {journal}
  {\bibinfo  {journal} {Phys. Rev. B},\ }\textbf {\bibinfo {volume} {79}},\
  \bibinfo {eid} {224302} (\bibinfo {year} {2009}{\natexlab{b}})}\BibitemShut
  {NoStop}%
\bibitem [{\citenamefont {Ray}\ and\ \citenamefont {Hallock}(2010)}]{Ray2010}%
  \BibitemOpen
  \bibfield  {author} {\bibinfo {author} {\bibfnamefont {M.}~\bibnamefont
  {Ray}}\ and\ \bibinfo {author} {\bibfnamefont {R.}~\bibnamefont {Hallock}},\
  }\Doi {10.1007/s10909-009-9975-y} {\bibfield  {journal} {\bibinfo  {journal}
  {J. Low Temp. Phys.},\ }\textbf {\bibinfo {volume} {158}},\ \bibinfo {pages}
  {560} (\bibinfo {year} {2010})}\BibitemShut {NoStop}%
\bibitem [{\citenamefont {Beamish}\ \emph {et~al.}(1983)\citenamefont
  {Beamish}, \citenamefont {Hikata}, \citenamefont {Tell},\ and\ \citenamefont
  {Elbaum}}]{Beamish1983}%
  \BibitemOpen
  \bibfield  {author} {\bibinfo {author} {\bibfnamefont {J.~R.}\ \bibnamefont
  {Beamish}}, \bibinfo {author} {\bibfnamefont {A.}~\bibnamefont {Hikata}},
  \bibinfo {author} {\bibfnamefont {L.}~\bibnamefont {Tell}}, \ and\ \bibinfo
  {author} {\bibfnamefont {C.}~\bibnamefont {Elbaum}},\ }\Doi
  {10.1103/PhysRevLett.50.425} {\bibfield  {journal} {\bibinfo  {journal}
  {Phys. Rev. Lett.},\ }\textbf {\bibinfo {volume} {50}},\ \bibinfo {pages}
  {425} (\bibinfo {year} {1983})}\BibitemShut {NoStop}%
\bibitem [{\citenamefont {Lie-zhao}\ \emph {et~al.}(1986)\citenamefont
  {Lie-zhao}, \citenamefont {Brewer}, \citenamefont {Girit}, \citenamefont
  {Smith},\ and\ \citenamefont {Reppy}}]{Lie-zhao1986}%
  \BibitemOpen
  \bibfield  {author} {\bibinfo {author} {\bibfnamefont {C.}~\bibnamefont
  {Lie-zhao}}, \bibinfo {author} {\bibfnamefont {D.~F.}\ \bibnamefont
  {Brewer}}, \bibinfo {author} {\bibfnamefont {C.}~\bibnamefont {Girit}},
  \bibinfo {author} {\bibfnamefont {E.~N.}\ \bibnamefont {Smith}}, \ and\
  \bibinfo {author} {\bibfnamefont {J.~D.}\ \bibnamefont {Reppy}},\ }\Doi
  {10.1103/PhysRevB.33.106} {\bibfield  {journal} {\bibinfo  {journal} {Phys.
  Rev. B},\ }\textbf {\bibinfo {volume} {33}},\ \bibinfo {pages} {106}
  (\bibinfo {year} {1986})}\BibitemShut {NoStop}%
\bibitem [{\citenamefont {Adams}\ \emph {et~al.}(1987)\citenamefont {Adams},
  \citenamefont {Tang}, \citenamefont {Uhlig},\ and\ \citenamefont
  {Haas}}]{Adams1987}%
  \BibitemOpen
  \bibfield  {author} {\bibinfo {author} {\bibfnamefont {E.}~\bibnamefont
  {Adams}}, \bibinfo {author} {\bibfnamefont {Y.}~\bibnamefont {Tang}},
  \bibinfo {author} {\bibfnamefont {K.}~\bibnamefont {Uhlig}}, \ and\ \bibinfo
  {author} {\bibfnamefont {G.}~\bibnamefont {Haas}},\ }\Doi
  {10.1007/BF00681469} {\bibfield  {journal} {\bibinfo  {journal} {J. Low Temp.
  Phys.},\ }\textbf {\bibinfo {volume} {66}},\ \bibinfo {pages} {85} (\bibinfo
  {year} {1987})}\BibitemShut {NoStop}%
\bibitem [{\citenamefont {\c{S} G.~S\"{o}yler}\ \emph
  {et~al.}(2009)\citenamefont {\c{S} G.~S\"{o}yler}, \citenamefont {Kuklov},
  \citenamefont {Pollet}, \citenamefont {Prokof'ev},\ and\ \citenamefont
  {Svistunov}}]{Soyler2009}%
  \BibitemOpen
  \bibfield  {author} {\bibinfo {author} {\bibnamefont {\c{S} G.~S\"{o}yler}},
  \bibinfo {author} {\bibfnamefont {A.~B.}\ \bibnamefont {Kuklov}}, \bibinfo
  {author} {\bibfnamefont {L.}~\bibnamefont {Pollet}}, \bibinfo {author}
  {\bibfnamefont {N.~V.}\ \bibnamefont {Prokof'ev}}, \ and\ \bibinfo {author}
  {\bibfnamefont {B.~V.}\ \bibnamefont {Svistunov}},\ }\Doi
  {10.1103/PhysRevLett.103.175301} {\bibfield  {journal} {\bibinfo  {journal}
  {Phys. Rev. Lett.},\ }\textbf {\bibinfo {volume} {103}},\ \bibinfo {eid}
  {175301} (\bibinfo {year} {2009})}\BibitemShut {NoStop}%
\bibitem [{\citenamefont {Straty}\ and\ \citenamefont
  {Adams}(1969)}]{Straty1969}%
  \BibitemOpen
  \bibfield  {author} {\bibinfo {author} {\bibfnamefont {G.~C.}\ \bibnamefont
  {Straty}}\ and\ \bibinfo {author} {\bibfnamefont {E.~E.}\ \bibnamefont
  {Adams}},\ }\href@noop {} {\bibfield  {journal} {\bibinfo  {journal} {Rev.
  Sci Inst.},\ }\textbf {\bibinfo {volume} {40}},\ \bibinfo {pages} {1393}
  (\bibinfo {year} {1969})}\BibitemShut {NoStop}%
\bibitem [{\citenamefont {Mikhin}\ \emph {et~al.}(2001)\citenamefont {Mikhin},
  \citenamefont {Polev},\ and\ \citenamefont {Rudavskii}}]{Mikhin2001}%
  \BibitemOpen
  \bibfield  {author} {\bibinfo {author} {\bibfnamefont {N.~P.}\ \bibnamefont
  {Mikhin}}, \bibinfo {author} {\bibfnamefont {A.~V.}\ \bibnamefont {Polev}}, \
  and\ \bibinfo {author} {\bibfnamefont {E.~Y.}\ \bibnamefont {Rudavskii}},\
  }\href@noop {} {\bibfield  {journal} {\bibinfo  {journal} {JETP Letters},\
  }\textbf {\bibinfo {volume} {73}},\ \bibinfo {pages} {470} (\bibinfo {year}
  {2001})},\ ISSN \bibinfo {issn} {00213640}\BibitemShut {NoStop}%
\bibitem [{\citenamefont {Mikhin}\ \emph {et~al.}(2007)\citenamefont {Mikhin},
  \citenamefont {Polev}, \citenamefont {Rudavskii},\ and\ \citenamefont
  {Vekhov}}]{Mikhin2007}%
  \BibitemOpen
  \bibfield  {author} {\bibinfo {author} {\bibfnamefont {N.}~\bibnamefont
  {Mikhin}}, \bibinfo {author} {\bibfnamefont {A.}~\bibnamefont {Polev}},
  \bibinfo {author} {\bibfnamefont {E.}~\bibnamefont {Rudavskii}}, \ and\
  \bibinfo {author} {\bibfnamefont {Y.}~\bibnamefont {Vekhov}},\ }\Doi
  {10.1007/s10909-007-9458-y} {\bibfield  {journal} {\bibinfo  {journal} {J.
  Low Temp. Phys.},\ }\textbf {\bibinfo {volume} {148}},\ \bibinfo {pages}
  {707} (\bibinfo {year} {2007})}\BibitemShut {NoStop}%
\bibitem [{\citenamefont {Pantalei}\ \emph {et~al.}(2010)\citenamefont
  {Pantalei}, \citenamefont {Rojas}, \citenamefont {Edwards}, \citenamefont
  {Maris},\ and\ \citenamefont {Balibar}}]{Pantalei2010}%
  \BibitemOpen
  \bibfield  {author} {\bibinfo {author} {\bibfnamefont {C.}~\bibnamefont
  {Pantalei}}, \bibinfo {author} {\bibfnamefont {X.}~\bibnamefont {Rojas}},
  \bibinfo {author} {\bibfnamefont {D.~O.}\ \bibnamefont {Edwards}}, \bibinfo
  {author} {\bibfnamefont {H.~J.}\ \bibnamefont {Maris}}, \ and\ \bibinfo
  {author} {\bibfnamefont {S.}~\bibnamefont {Balibar}},\ }\Doi
  {10.1007/s10909-010-0159-6} {\bibfield  {journal} {\bibinfo  {journal} {J.
  Low Temp. Phys.},\ }\textbf {\bibinfo {volume} {159}},\ \bibinfo {pages}
  {452} (\bibinfo {year} {2010})}\BibitemShut {NoStop}%
\bibitem [{\citenamefont {Swenson}(1950)}]{Swenson1950}%
  \BibitemOpen
  \bibfield  {author} {\bibinfo {author} {\bibfnamefont {C.~A.}\ \bibnamefont
  {Swenson}},\ }\Doi {10.1103/PhysRev.79.626} {\bibfield  {journal} {\bibinfo
  {journal} {Phys. Rev.},\ }\textbf {\bibinfo {volume} {79}},\ \bibinfo {pages}
  {626} (\bibinfo {year} {1950})}\BibitemShut {NoStop}%
\bibitem [{\citenamefont {Hiroi}\ \emph {et~al.}(1989)\citenamefont {Hiroi},
  \citenamefont {Mizusaki}, \citenamefont {Tsuneto}, \citenamefont {Hirai},\
  and\ \citenamefont {Eguchi}}]{Hiroi1989}%
  \BibitemOpen
  \bibfield  {author} {\bibinfo {author} {\bibfnamefont {M.}~\bibnamefont
  {Hiroi}}, \bibinfo {author} {\bibfnamefont {T.}~\bibnamefont {Mizusaki}},
  \bibinfo {author} {\bibfnamefont {T.}~\bibnamefont {Tsuneto}}, \bibinfo
  {author} {\bibfnamefont {A.}~\bibnamefont {Hirai}}, \ and\ \bibinfo {author}
  {\bibfnamefont {K.}~\bibnamefont {Eguchi}},\ }\Doi {10.1103/PhysRevB.40.6581}
  {\bibfield  {journal} {\bibinfo  {journal} {Phys. Rev. B},\ }\textbf
  {\bibinfo {volume} {40}},\ \bibinfo {pages} {6581} (\bibinfo {year}
  {1989})}\BibitemShut {NoStop}%
\end{thebibliography}%

\end{document}